\documentclass[journal]{IEEEtran}

\usepackage{amsmath,amssymb,amsfonts,mathtools}

\usepackage[utf8]{inputenc}
\usepackage[T1]{fontenc}
\usepackage{hyperref}
\usepackage{url}
\usepackage{booktabs}
\usepackage{array}
\usepackage{microtype}
\usepackage{xcolor}
\usepackage{graphicx}
\usepackage{subcaption}
\usepackage{multirow}
\usepackage{cite}

\usepackage[capitalize,noabbrev]{cleveref}

\title{An Overview on Multimodal Wireless and Visual Sensing for Environment Reconstruction: Opportunities and Challenges}

\author{Guanlin Wu, Chao Hu, Yilong Chen, Yudong Guo, Rongfei Fan, Juyong Zhang, Han Hu,\\ and Jie Xu, \textit{Fellow, IEEE}%
    \thanks{G. Wu, C. Hu, Y. Chen, and J. Xu are with the School of Science and Engineering, the Shenzhen Future Network of Intelligence Institute, and the Guangdong Provincial Key Laboratory of Future Networks of Intelligence, The Chinese University of Hong Kong, Shenzhen, Guangdong 518172, China (e-mail: guanlinwu1@link.cuhk.edu.cn, chaohu@link.cuhk.edu.cn, yilongchen@link.cuhk.edu.cn, xujie@cuhk.edu.cn). J. Xu is the corresponding author.}
    \thanks{Y. Guo and J. Zhang are with the School of Mathematical Sciences, University of Science and Technology of China, Hefei, Anhui 230026, China (e-mail: yudong@ustc.edu.cn, juyong@ustc.edu.cn).}
    \thanks{R. Fan is with the School of Cyberspace Science and Technology, Beijing Institute of Technology, Beijing 100081, China (e-mail: fanrongfei@bit.edu.cn).}
    \thanks{H. Hu is with the School of Information and Electronics, Beijing Institute of Technology, Beijing 100081, China (e-mail: hhu@bit.edu.cn).}
    }

\begin{document}

\maketitle

\begin{abstract}
Physical-scene reconstruction through visual sensing and channel knowledge map (CKM) reconstruction through wireless sensing have attracted growing research interest but have largely evolved independently within the computer vision and wireless communications communities, respectively. When applied to the same physical environment, the two sensing modalities provide complementary observations. Visual sensing captures detailed scene geometry and appearance but is constrained by occlusion and poor lighting. Wireless sensing can provide information beyond visual line of sight regardless of lighting conditions, but its measurements are often sparse and difficult to associate with scene structures. These complementary strengths motivate a unified multimodal approach to physical-scene and CKM reconstruction. This paper provides an overview of such unified environment reconstruction through wireless and visual sensing. We first compare the capabilities and limitations of the two modalities and discuss potential applications that benefit from jointly reconstructing both targets. We then review two complementary research directions, namely wireless-assisted physical-scene reconstruction and vision-assisted CKM reconstruction, highlighting how each modality supports the reconstruction task associated with the other. Building on these perspectives, we discuss design principles for a joint representation that supports both targets and identify mutual refinement, in which each modality helps improve the reconstruction associated with the other. Finally, we discuss open challenges and future research directions, offering insights into developing site-specific environment models for sixth-generation (6G) and beyond networks.
\end{abstract}


\section{Introduction}
\label{sec:intro}

Physical-scene reconstruction via visual sensing has attracted considerable research interest in computer vision and graphics, with applications in autonomous navigation and extended reality, as well as digital twins~\cite{christodoulides2025survey3d,11355839}. It aims to recover the geometry and appearance of a physical environment, including its spatial layout, from visual observations. Cameras provide images of visible surfaces, while red-green-blue-depth (RGB-D) sensors and light detection and ranging (LiDAR) provide metric depth measurements or 3D point clouds. Existing methods represent the reconstructed geometry explicitly through point clouds and meshes~\cite{curless1996volumetric,kazhdan2006poisson}, or implicitly through occupancy and signed-distance functions~\cite{mescheder2019occupancy,park2019deepsdf}. More recently, neural radiance fields (NeRF) have modeled scene density and view-dependent appearance using continuous neural functions, while 3D Gaussian splatting (3DGS) represents scenes through Gaussian primitives for efficient rendering of novel views~\cite{mildenhall2020nerf,kerbl2023gaussian}. Despite these advances, reconstruction accuracy remains limited by the available observations. Images provide little direct information about occluded surfaces, and monocular depth estimation cannot determine absolute scene scale without additional constraints. Although RGB-D sensors and LiDAR provide metric measurements, their coverage remains limited by occlusion and sensing range.

On the other hand, channel knowledge map (CKM) reconstruction via wireless sensing has attracted growing interest in the wireless communications networks to provide site-specific channel information for environment-aware network operation~\cite{zeng2021ckm,zeng2024ckmtutorial,ren2025ckmconstruction,li2023ckmuav,wang2023radiomapprecoding,gong2023radiomapuav,wang2023road6g,liu2022isac,liu2026cooperativeisac,zhang2026weit}. A CKM associates wireless channel quantities, such as received power and multipath parameters, with spatial locations of transmitter-receiver links. CKM reconstruction aims to estimate these quantities at transceiver locations where direct channel measurements are unavailable. Wireless sensing (like radar and integrated sensing and communications (ISAC)) provides the required propagation information through channel measurements. Existing approaches construct CKMs using measured or simulated channel data, often together with physical-scene information. For instance, ray tracing computes propagation paths and channel responses from site-specific geometry and the electromagnetic properties of surfaces~\cite{hoydis2023sionnaRT}. Learning-based methods predict channel quantities from environmental information and link configurations using simulated or measured labels~\cite{ckmimagenet2025}. Inspired by radiance-field representations in computer vision, wireless radiance fields learn continuous spatial representations of radio propagation from radio-frequency (RF) measurements~\cite{zhao2023nerf2,wrfgs2024,zhou2025biwgs}. These approaches support channel prediction beyond directly measured links, but their accuracy depends on the available RF information and how well the model captures the actual propagation environment. 

While visual and wireless sensing have been applied for physical-scene and CKM reconstruction separately, they provide different types of information about the environment. Visual sensing captures detailed geometric and appearance information from observed surfaces, but its effectiveness depends on visibility and sensing conditions. In particular, occlusion and poor illumination can leave parts of the scene insufficiently observed. By contrast, wireless sensing measures signal propagation through the environment and can remain informative beyond the optical line of sight (LoS), depending on the signal frequency and the properties of intervening materials. More specifically, radar measurements can provide range and motion information, while communication signals reveal propagation conditions along the measured links. However, wireless observations are typically sparse, and multipath propagation makes it difficult to associate individual measurements with specific scene structures. Therefore, the two modalities of visual and wireless sensing provide complementary information for the same environment. Wireless measurements can constrain geometry where visual observations are unreliable, while visual information can identify the environmental structures responsible for blockage and reflected propagation paths.

\begin{figure*}[!t]
    \centering
    \includegraphics[width=0.98\textwidth]{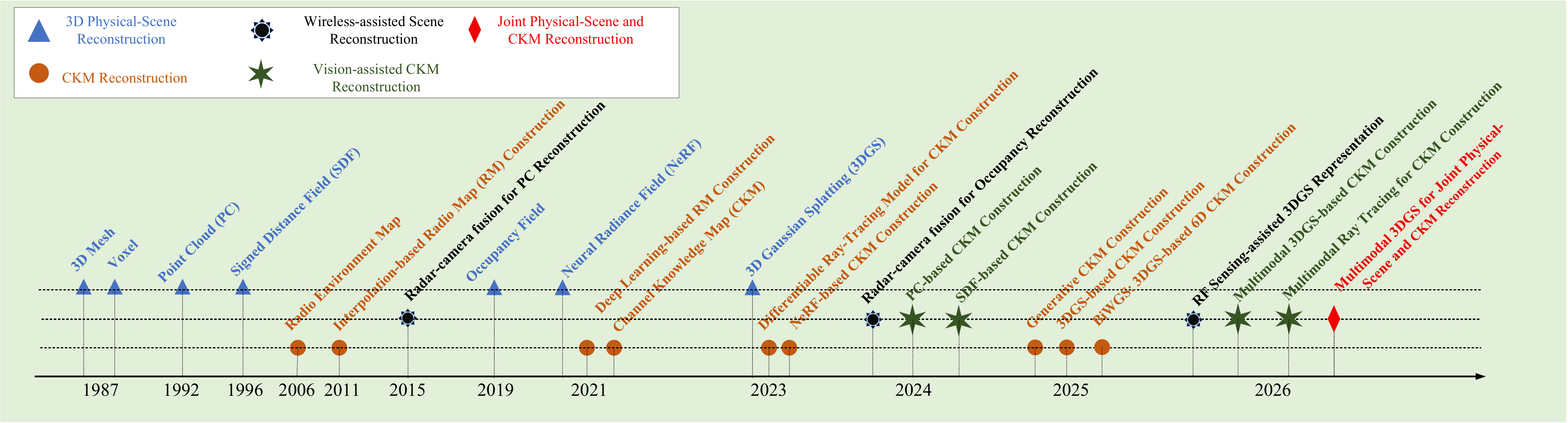}
    \caption{Milestones in multimodal wireless and visual sensing for environment reconstruction.}
    \label{fig:milestone}
\end{figure*}

Furthermore, the connection between physical-scene and CKM reconstruction extends beyond their complementary sensing modalities. Both tasks describe the same physical environment, although they recover different properties of it. Scene geometry determines which propagation paths can exist between a transmitter and a receiver, while the electromagnetic properties of the interacting surfaces determine how radio signals are reflected or attenuated. Consequently, geometric information recovered from visual sensing can constrain CKM reconstruction, and wireless measurements can provide additional constraints on uncertain scene geometry. Recent advances have also brought the two research areas closer at the representation level. Radiance-field and Gaussian-based models, initially developed for visual reconstruction, have been extended to represent wireless propagation. For example, URF-GS uses a unified Gaussian representation to support optical rendering and radio prediction~\cite{urfgs2026}. RadarSim models radar observations using a scene field initialized from visual information while allowing radar-specific geometric deviations~\cite{chen2026radarsim}, and NeuRadar combines a shared scene representation with a radar-specific observation model~\cite{rafidashti2025neuradar}. These developments motivate unified multimodal environment reconstruction, in which visual and wireless observations support both reconstruction targets. They also raise the possibility of mutual refinement, where wireless measurements improve the reconstructed physical scene and the updated scene, in turn, improves CKM reconstruction. Figure~\ref{fig:milestone} summarizes representative milestones in the development of these research directions.

Existing surveys have reviewed different aspects of environment reconstruction, primarily from their respective research communities. Surveys of 3D scene reconstruction examine methods ranging from multi-view geometry to neural scene representations, with an emphasis on recovering geometric structure and visual appearance~\cite{christodoulides2025survey3d,11355839,liu2025survey3dreconstruction,wu2024multiview3dreconstruction,wu2024recent3dgs,zhang2025feedforward3dsurvey}. Surveys and tutorials on radio maps and CKMs focus on estimating spatial channel information from measurements or propagation models~\cite{romero2022radiomap,zeng2024ckmtutorial,radiomaptutorial2026,ren2025ckmconstruction}. Studies reviewing millimeter-wave (mmWave) radar and multimodal sensing mainly address perception tasks through the combination of radar with cameras or LiDAR~\cite{wang2024mmwavefusion,kong2025mmwavesurvey}. These surveys provide important insights into their respective tasks, but they do not systematically examine how wireless observations contribute to physical-scene reconstruction or how visual information supports CKM reconstruction within a common framework. In particular, the different reconstruction targets and evaluation criteria make it difficult to compare the benefits of cross-modal sensing. The emerging convergence of their representations further motivates a unified examination of the two tasks and the opportunities for joint reconstruction.

This paper provides an overview of multimodal wireless and visual sensing for environment reconstruction, covering both physical-scene and CKM reconstruction. Section~\ref{sec:framework} examines the capabilities and limitations of the two sensing modalities, introduces the main paradigms of multimodal reconstruction, and discusses representative application scenarios. Section~\ref{sec:env} reviews wireless-assisted physical-scene reconstruction, organizing existing methods according to their scene representations and examining how wireless measurements improve reconstruction under limited visual observations. Section~\ref{sec:chan} discusses vision-assisted CKM reconstruction through ray tracing and learning-based approaches, as well as wireless radiance fields, with emphasis on how visual information supports channel prediction. Section~\ref{sec:bridge} examines joint reconstruction and the principles of unified environment representations that can support mutual refinement between the two tasks. Section~\ref{sec:future} discusses the remaining challenges in multimodal environment reconstruction and identifies promising future research directions. Section~\ref{sec:conclusion} finally concludes the paper.

\section{Multimodal Wireless and Visual Sensing for Environment Reconstruction}
\label{sec:framework}
\label{subsec:single_modality_limitations}

Physical-scene reconstruction and CKM recover different properties of the same environment. The former describes scene geometry and appearance, while the latter characterizes radio propagation across locations and communication links. Their reconstruction accuracy is evaluated against different references: physical-scene reconstruction uses geometric or visual measurements, whereas CKM reconstruction uses channel measurements at different transceiver locations. Although the two tasks have different outputs, both depend on the physical structures of the site. This section examines the properties of visual and wireless sensing, introduces three modes of multimodal environment reconstruction, and discusses representative applications.

\subsection{Visual Sensing}
\label{subsec:visual_sensing}

Visual sensing acquires information about the environment through optical observations. Cameras record light reflected from visible surfaces, providing detailed appearance and spatial information from which scene geometry can be estimated. Depth sensors and LiDAR provide additional distance measurements, allowing observed surfaces to be located in metric 3D space. These observations support the reconstruction of scene structures and their spatial relationships. Camera images offer dense spatial information, while depth measurements help recover the geometric scale that cannot be determined from a single image alone.

The reliability of visual sensing depends on the visibility and optical properties of the observed surfaces. Objects outside the sensor's field of view or behind other objects cannot be directly observed, and camera-based reconstruction is sensitive to illumination conditions~\cite{cheng2024som}. Although LiDAR provides metric range measurements, its returns can degrade in adverse weather, such as rain or fog~\cite{wang2024mmwavefusion}. Visual reconstruction also faces geometric ambiguity. Monocular images do not directly determine absolute depth scale, and depth estimation becomes less reliable on surfaces with weak texture~\cite{cheng2024som,javadi2026singleimage}. Reflective or transparent surfaces can further produce inaccurate or missing depth measurements~\cite{christodoulides2025survey3d}. Consequently, the reconstructed scene may contain uncertain geometry in regions with insufficient or unreliable optical observations.

\subsection{Wireless Sensing}
\label{subsec:wireless_sensing}

Wireless sensing acquires environmental information from the propagation of RF signals. Radar and ISAC systems process received echoes to estimate the range and direction of reflecting objects, as well as their radial motion through Doppler shifts. Communication receivers instead observe the responses of active wireless links, including received power and channel state information (CSI). When individual multipath components can be resolved, their delays and arrival directions provide additional information about the structures encountered by the signals. Unlike camera observations, RF measurements do not depend on ambient illumination. Radio signals can also remain informative beyond optical LoS, although penetration through obstacles depends on signal frequency and material properties~\cite{wang2024mmwavefusion}.

Wireless observations have different limitations from their visual counterparts. Radar measurements generally provide lower angular resolution than camera images, and clutter or multipath propagation can make it difficult to associate individual echoes with physical objects~\cite{wang2024mmwavefusion,kong2025mmwavesurvey}. Communication-channel measurements are available only for the transmitter-receiver configurations being observed. Predicting the channel at unmeasured locations therefore requires additional information about the environment, particularly when blockage causes abrupt spatial variations~\cite{ren2025ckmconstruction}. Moreover, received power and aggregate CSI combine contributions from multiple propagation paths, so different environmental structures may produce similar channel observations. Path-resolved measurements can reduce this ambiguity, but their resolution remains limited by the available bandwidth and antenna aperture~\cite{zeng2024ckmtutorial,wang2024mmwavefusion}. As a result, wireless measurements provide direct information about propagation but cannot always determine the corresponding scene structures.

\subsection{Multimodal Environment Reconstruction}
\label{subsec:two_directions}

The two sensing modalities provide complementary observations because they interact with the same physical environment through different mechanisms. Visual sensing identifies the geometry of visible surfaces, while wireless measurements provide range or propagation information that may remain available where optical observations are unreliable. The relationship also works in the opposite direction. A reconstructed wall can indicate whether a direct propagation path is blocked, and its orientation helps determine possible reflection paths. However, the strength of these paths depends on the electromagnetic properties of the interacting surfaces, as well as the transmitter-receiver configuration. Visual geometry therefore constrains possible propagation paths, while RF measurements or propagation models determine their channel responses.

These relationships give rise to three modes of multimodal environment reconstruction. They differ in which reconstruction target is considered and how observations from the other modality are incorporated. The first two modes use one sensing modality to improve the reconstruction associated with the other, while the third considers the joint reconstruction of physical-scene and CKM.

\subsubsection{Wireless-Assisted Physical-Scene Reconstruction}

Wireless-assisted physical-scene reconstruction incorporates RF observations as an additional modality into the estimation of scene geometry or motion. For example, a radar measurement provides the distance to a reflecting object. Once the return is associated with a surface observed by a camera, the measured range can constrain the surface position and reduce the scale ambiguity of monocular depth. Radar returns can also provide information about objects that are poorly observed under low illumination or partial occlusion. These measurements improve the robustness of physical-scene reconstruction by supplying geometric constraints where visual information is insufficient.

However, RF measurements are typically sparse and do not directly describe complete object surfaces. Their contribution depends on associating the measurements with the correct scene structures and extending the resulting constraints to regions without RF observations. Existing approaches address these problems using different scene representations, as reviewed in Section~\ref{sec:env}.

\subsubsection{Vision-Assisted CKM Reconstruction}

Vision-assisted CKM reconstruction uses visually recovered environmental information to support the estimation of wireless channel quantities. Reconstructed scene geometry identifies obstacles and reflecting surfaces, allowing propagation paths to be evaluated for different transmitter-receiver configurations. For example, a building reconstructed from camera images can be used to determine which links are blocked and which surfaces may support reflected paths. The corresponding channel gains still depend on electromagnetic properties that must be specified through a propagation model or estimated from RF measurements.

By relating channel observations to the reconstructed environment, visual information helps predict wireless propagation at locations where direct measurements are unavailable. It can also reduce the amount of RF data needed to construct a CKM when the relevant scene geometry is sufficiently accurate. The remaining problem is to determine the channel response from the available geometry and RF information, especially when the electromagnetic properties of surfaces are uncertain. Section~\ref{sec:chan} reviews approaches based on ray tracing and learning, as well as wireless radiance fields.

\subsubsection{Joint Physical-Scene and CKM Reconstruction}

Joint reconstruction considers both the physical scene and its wireless propagation characteristics within the same environment. The two reconstruction targets are related through scene geometry, but they require different response models. A physical surface has a visual appearance and an electromagnetic response, and the two cannot generally be inferred from each other. Joint reconstruction can therefore exploit common environmental information while retaining the properties needed to predict each type of observation.

A unified environment representation offers one possible way to support both targets. Such a representation may share scene geometry while separately modeling optical and RF responses. Beyond producing both outputs, mutual refinement would allow wireless measurements to correct uncertain scene geometry and the updated geometry to improve CKM reconstruction. This requires the two modalities to be aligned in space and time, together with observation models that distinguish geometric errors from errors in the predicted sensor responses. Section~\ref{sec:bridge} discusses the principles and remaining challenges of this joint setting.

\subsection{Representative Applications}
\label{subsec:applications}

\begin{figure}[!t]
    \centering
    \includegraphics[width=\columnwidth]{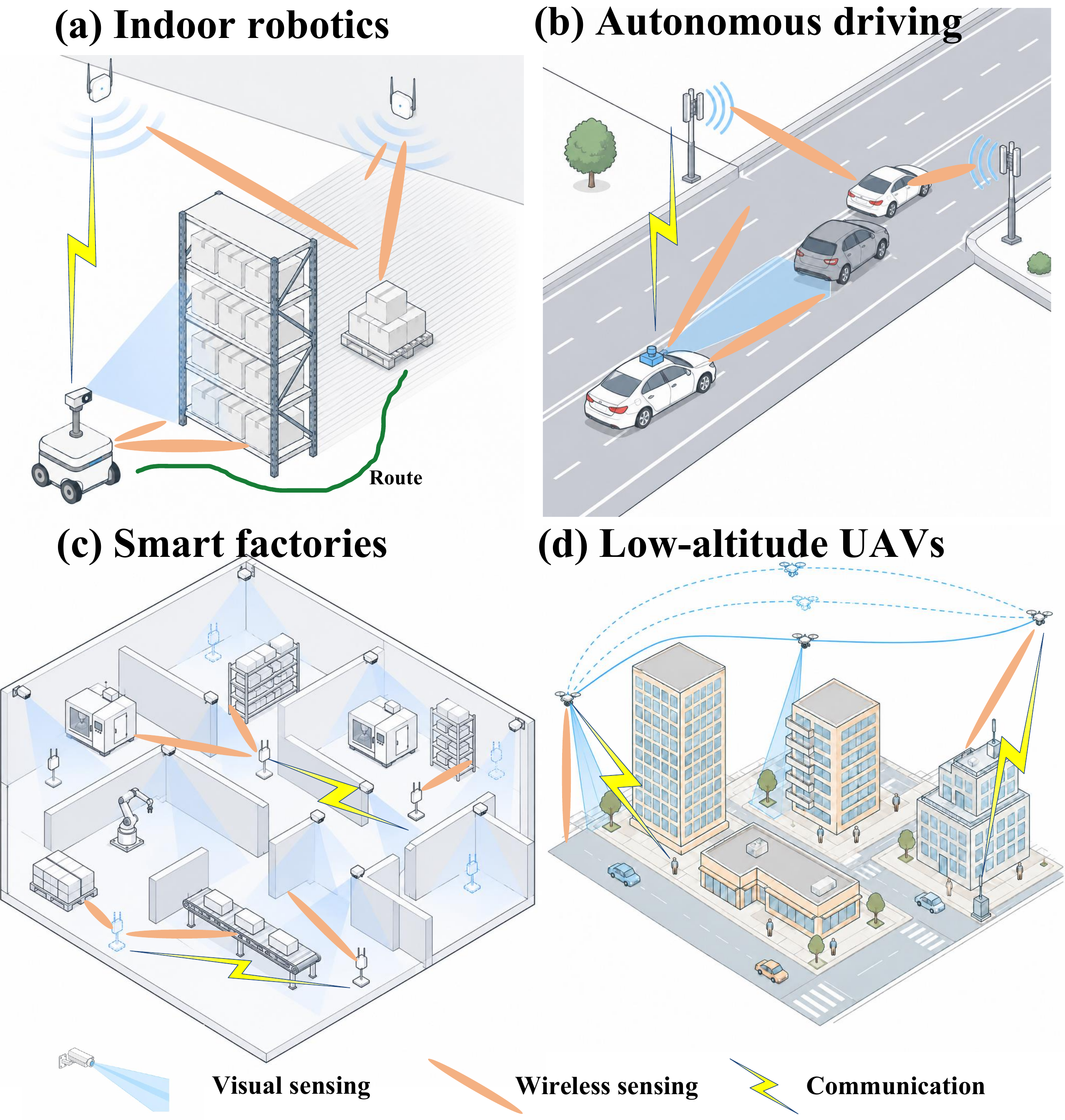}
    \caption{Multimodal wireless and visual sensing in four representative applications. (a) Indoor robotics: the robot and access points (APs) jointly sense the shelving to support route planning. (b) Autonomous driving: the vehicle and roadside units (RSUs) combine visual and wireless sensing to perceive the surrounding traffic. (c) Smart factories: cameras and APs jointly sense the environment topology to guide AP deployment, with dashed APs marking candidate positions. (d) Low-altitude UAVs: UAVs fuse visual and wireless sensing to plan flight corridors that support air-to-ground communication; the solid line shows the selected path, and dashed lines show alternatives.}
    \label{fig:application_scenarios}
\end{figure}

Several applications require both physical-scene information and site-specific wireless channel knowledge. Figure~\ref{fig:application_scenarios} illustrates four representative scenarios in which the same environmental structures affect physical operations and wireless connectivity. The two reconstruction targets provide different information for these applications, while their common geometric basis creates opportunities for multimodal sensing.

\emph{Indoor robotics:} A mobile robot operating in a warehouse needs an accurate scene model for navigation and reliable wireless connectivity for coordination with other robots. Visual mapping can be unreliable in long aisles with repetitive shelving or uneven lighting, while monocular depth estimates may have uncertain metric scale. Radar measurements can improve monocular depth estimation under these conditions~\cite{job2024radarmeetsvision}. Wi-Fi APs can also provide spatial references that reduce accumulated localization errors in simultaneous localization and mapping (SLAM)~\cite{arun2024wais}. For wireless connectivity, the same reconstructed shelving and walls determine which propagation paths may be blocked or reflected along a robot's route. A scene model that also provides predicted received power can therefore support route planning under both obstacle-avoidance and connectivity requirements, as demonstrated in an early unified reconstruction approach~\cite{urfgs2026}.

\emph{Autonomous driving:} A connected vehicle must identify obstacles and drivable space while maintaining communication with RSUs. The same surrounding objects can affect both functions. For example, a truck may obstruct the vehicle's camera view and block its LoS wireless link to an RSU. Radar measurements complement cameras under poor illumination or adverse weather, improving semantic occupancy prediction~\cite{ming2025occfusion}. Visual information also helps anticipate changes in wireless connectivity. Camera images have been used to predict impending link blockage and initiate proactive handoff~\cite{charan2021visionaided}, while LiDAR point clouds support future beam prediction in measured vehicle-to-infrastructure links~\cite{jiang2023lidarbeam}. Combining LiDAR observations with received pilot signals can further improve CSI estimation with fewer pilots~\cite{zhang2025somcsi}. These techniques illustrate how scene geometry can support wireless prediction and link management, while RF sensing contributes to scene perception.

\emph{Smart factories:} A smart factory requires an accurate representation of its equipment layout to support physical operations and wireless network planning. Industrial environments contain machinery and other structures that may block or reflect radio signals, and changes in equipment placement can alter wireless coverage. Reconstructing the factory from visual observations provides the geometry needed for site-specific channel prediction. For example, HoRAMA reconstructs a factory environment from smartphone video, assigns material properties, and validates the resulting ray-tracing model against channel soundings~\cite{ying2026horama}. Measured channel data can further calibrate the electromagnetic parameters used in industrial ray tracing~\cite{gougeon2024industrialcalibration}, allowing the reconstructed model to support AP placement and coverage planning~\cite{liao2022haru}. When equipment moves, visual observations can identify changes in scene geometry, although the resulting channel response must still be estimated or measured. RFCanvas uses visually estimated motion to update its RF field model, reducing the median channel prediction error in dynamic scenes from 3.5~dB to 2.2~dB~\cite{rfcanvas2024}.

\emph{Low-altitude unmanned aerial vehicles:} A low-altitude unmanned aerial vehicle (UAV) requires a 3D scene model for navigation and site-specific channel information for air-to-ground communication. Buildings affect both tasks because they provide geometric references for localization while also blocking or reflecting wireless signals. Near building facades, satellite positioning can become unreliable, making accurate local 3D reconstruction particularly important. Sparse mmWave radar measurements have been used to refine visual point clouds for map-based UAV localization~\cite{cui2026uavlocalization}. Wireless coverage also varies with the UAV's altitude and position, so measurements collected along one flight trajectory provide limited information about other locations. By combining onboard depth observations with received signal strength (RSS) measurements, existing work has jointly estimated building geometry and wireless channel parameters in simulation, allowing radio maps to be predicted at UAV positions where no measurements were collected~\cite{esrafilian2021mapfusion}.

\section{Wireless-Assisted Physical-Scene Reconstruction}
\label{sec:env}

Wireless-assisted physical-scene reconstruction incorporates RF observations into visual reconstruction to improve the estimation of scene geometry and appearance. Radar echoes provide metric range and radial velocity information, while multipath components of communication signals can reveal the distances and directions of propagation paths interacting with the environment. These measurements provide additional constraints where visual observations are uncertain. However, RF observations are typically sparse, and their geometric interpretation depends on associating the measured signals with physical structures. Existing methods therefore differ in how they incorporate RF information into the reconstructed scene. Based on the target representation, they can be organized into four categories: depth and point clouds, occupancy, surfaces and meshes, and NeRF and 3DGS. Table~\ref{tab:wireless_scene_recon_taxonomy} summarizes representative methods and the roles of wireless observations in their reconstruction processes.

\begin{table*}[t]
\centering
\caption{Representative wireless-assisted methods organized by scene representation.}
\label{tab:wireless_scene_recon_taxonomy}
\footnotesize
\setlength{\tabcolsep}{3pt}
\renewcommand{\arraystretch}{1.18}
\begin{tabular}{@{}
>{\raggedright\arraybackslash}m{0.14\linewidth}
>{\raggedright\arraybackslash}m{0.19\linewidth}
>{\raggedright\arraybackslash}m{0.16\linewidth}
>{\raggedright\arraybackslash}m{0.31\linewidth}
>{\raggedright\arraybackslash}m{0.14\linewidth}
@{}}
\toprule
\textbf{Scene Representation} &
\textbf{Representative Work} &
\textbf{Wireless Information} &
\textbf{Role in Reconstruction} &
\textbf{Output} \\
\midrule

Depth and point clouds &
Radar--vision 3D~\cite{elnatour2015toward3d} &
Radar range and bearing &
Associates radar measurements with image observations through calibrated sensor geometry to recover metric 3D points &
Textured 3D points \\

\cmidrule(l){2-5}

&
ISAC environment reconstruction~\cite{lin2024isacenvreconstruction} &
Multipath parameters of selected users &
Estimates user and reflection locations from propagation paths and combines them with image-derived depth &
Reflection points and dense depth \\

\cmidrule(l){2-5}

&
Radar-assisted novel-view synthesis~\cite{javadi2026singleimage} &
Sparse radar range samples &
Uses radar-constrained depth estimation to construct geometry for novel-view synthesis &
Dense depth and generated views \\

\cmidrule(l){2-5}

&
mmWave--vision semantic 3D~\cite{yang2023mmwavevisionsemantic3d} &
Radar return intensity &
Associates radar intensity with point-cloud geometry and image-derived semantic labels &
Semantic point cloud with intensity \\

\midrule

Occupancy &
OccFusion~\cite{ming2025occfusion} &
Radar point coordinates and velocity features &
Aligns camera and radar features in spatial grids and learns occupancy in cells without direct radar returns &
Semantic occupancy \\

\midrule

Surfaces and meshes &
ImmFusion~\cite{chen2023immfusion} &
mmWave radar points &
Combines image and radar features with a human-body template to estimate body shape &
Human joints and mesh \\

\midrule

NeRF and 3DGS &
Multimodal Gaussian splatting~\cite{gau2026multimodalgs} &
Sparse RF-derived depth &
Estimates dense depth from sparse range constraints to initialize Gaussian positions before image-based optimization &
Renderable Gaussian scene \\

\cmidrule(l){2-5}

&
Rad-GS~\cite{xiao2025radgs} &
4D radar points and Doppler velocities &
Uses Doppler to mask moving objects in images and radar geometry to track poses and constrain Gaussian depth &
Static Gaussian map \\

\bottomrule
\end{tabular}
\end{table*}

\subsection{Depth and Point-Cloud Reconstruction}
\label{sec:env:geometry_depth}

Depth maps describe the distance from a camera to the observed surfaces, while point clouds represent their positions in 3D space. Monocular images can provide dense depth estimates, but their absolute scale remains ambiguous without additional information. Depth estimates may also become unreliable under poor illumination or on surfaces with weak texture. Radar measurements provide metric distances to reflecting objects and can remain available under these conditions. Once associated with the corresponding visual observations, these measurements constrain the estimated surface positions and improve the recovered geometry.

Existing methods incorporate wireless range information through geometric association and depth estimation. For a calibrated radar-camera system, radar range and bearing measurements can be associated with image observations to recover metric 3D points~\cite{elnatour2015toward3d}. Communication signals can also provide geometric constraints when individual propagation paths are resolved. For example, an ISAC-based method estimates user locations from LoS paths and identifies reflection points from single-bounce non-LoS (NLoS) paths under the corresponding geometric assumptions~\cite{lin2024isacenvreconstruction}. These points are then combined with image-derived depth to improve the reconstructed scene. Since the RF measurements cover only a limited number of locations, their geometric information must also be extended to surrounding regions. Radar-assisted depth estimation addresses this problem by using sparse range samples to constrain local depth predictions while accounting for their uncertainty~\cite{javadi2026singleimage}.

The contribution of wireless observations depends on where the range constraints are available and how reliably they can be associated with the scene. Measurements near uncertain depth boundaries can help correct geometric discontinuities, while measurements on well-reconstructed surfaces may provide less additional information. Extending sparse constraints over large neighborhoods increases spatial coverage but can smooth important depth variations. Conversely, restricting their influence to small neighborhoods leaves more regions dependent on visual estimates. Communication-based reconstruction also depends on the resolution of individual propagation paths and the assumptions used to locate reflection points. Furthermore, radar return intensity varies with sensor pose and incidence angle, making its association with scene geometry different from that of visual appearance. Methods that attach radar intensity to image-derived semantic labels, such as~\cite{yang2023mmwavevisionsemantic3d}, must account for this observation dependence.

\begin{figure}[t]
    \centering
    \includegraphics[width=\columnwidth]{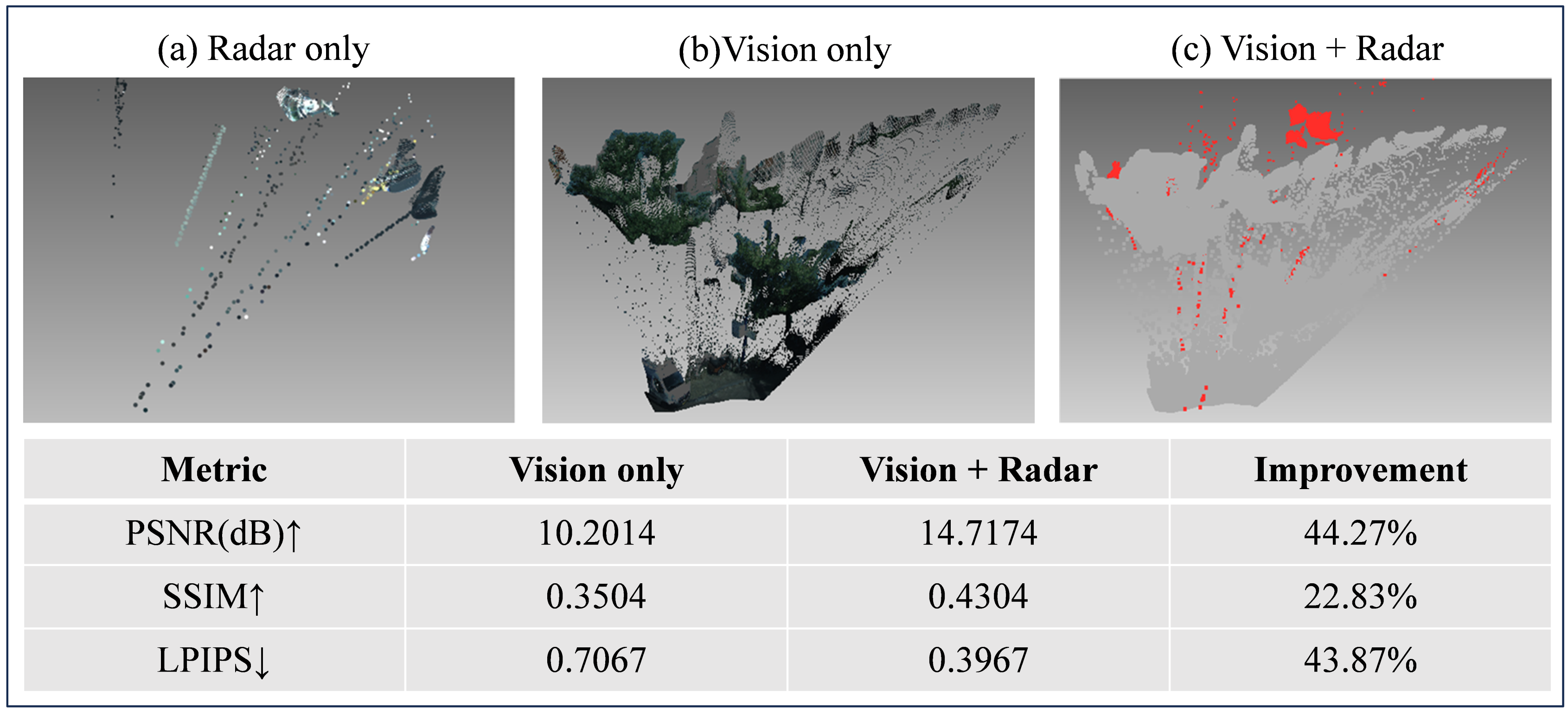}
    \caption{Example of radar-assisted point-cloud reconstruction for novel-view synthesis. The upper panels show point clouds obtained from (a) radar only, (b) vision only, and (c) vision and radar. The table compares the novel-view synthesis performance of vision-only and radar-assisted reconstruction. The radar-only point cloud is shown for qualitative comparison but is excluded from the quantitative evaluation due to its sparsity.}
    \label{fig:case_study_multimodal_pointcloud_nvs}
\end{figure}

\textbf{Example of Radar-Assisted Novel-View Synthesis.}
Figure~\ref{fig:case_study_multimodal_pointcloud_nvs} illustrates how sparse radar measurements can improve the geometry used for novel-view synthesis. Using the public implementation of~\cite{javadi2026singleimage}, the vision-only branch estimates monocular depth from an RGB image and back-projects it into a point cloud. The radar-assisted branch replaces this monocular depth with a dense depth map reconstructed from sparse metric range measurements before generating views from unobserved camera positions. The radar-only point cloud is much sparser than the image-derived reconstruction and is therefore shown separately to illustrate the available geometric measurements.
In the illustrated comparison, incorporating radar measurements increases the peak signal-to-noise ratio (PSNR) from 10.2014 to 14.7174~dB and the structural similarity index measure (SSIM) from 0.3504 to 0.4304. The learned perceptual image patch similarity (LPIPS) decreases from 0.7067 to 0.3967. These results show that sparse range constraints can improve the quality of generated views when they correct errors in the underlying visual geometry. The comparison also illustrates the different roles of the two modalities: visual observations provide dense scene information, while radar supplies metric constraints at selected locations.

\subsection{Occupancy Reconstruction}
\label{sec:env:occupancy}

Occupancy reconstruction estimates whether spatial cells are occupied by physical objects. Semantic occupancy further assigns object categories to the occupied cells, providing a volumetric description of the scene. Unlike a depth map, an occupancy representation can estimate the state of regions that are not directly visible from the current viewpoint. However, camera-based occupancy prediction becomes unreliable when objects are occluded or poorly illuminated. Radar measurements provide additional observations of reflecting structures and are less affected by illumination. Their spatial coverage is nevertheless sparse, and a radar return generally identifies only part of an object rather than its complete occupied volume.

Camera-radar fusion therefore requires the two modalities to contribute to occupancy estimates in a common spatial representation. OccFusion aligns image features and radar detections within spatial grids, allowing observations from both sensors to inform the occupancy of corresponding cells~\cite{ming2025occfusion}. The model then learns spatial relationships that propagate information from observed regions to cells without direct radar returns. Radar coordinates provide geometric evidence of occupied regions, while image features supply additional information about object boundaries and semantics. This combination improves semantic occupancy prediction under adverse weather and low-light conditions compared with camera-only estimation.

The main limitation is that missing radar detections do not establish that a region is empty. A surface may fail to produce a measurable return because of its material properties or orientation relative to the sensor. Occupancy in such regions must therefore be inferred from visual information or observations acquired from other viewpoints. Increasing the spatial resolution of the occupancy grid can represent smaller structures, but it does not increase the number of available radar measurements. As a result, finer occupancy predictions remain dependent on the model's ability to associate sparse detections with the surrounding scene.

\subsection{Surface and Mesh Reconstruction}
\label{sec:env:meshes}

Surface meshes represent object geometry through connected vertices and faces, providing an explicit description of shape that can be used for geometric processing and rendering. Reconstructing a complete mesh from visual observations is difficult when parts of the object are occluded. Radar can provide measurements from some regions that are not optically visible, since RF signals may penetrate certain non-metallic materials depending on frequency and material properties. However, radar returns depend on the electromagnetic response of the observed surfaces, making their positions difficult to infer from received signals alone. Combining visual and wireless observations can reduce this ambiguity by constraining the geometry that explains the radar measurements.

Existing methods use visual geometry or structural priors to recover surfaces from incomplete RF observations. For example, GeRaF 2.0 reconstructs the visible exterior of a container from camera images and uses this geometry to model radar propagation through its walls~\cite{lu2026seeingthroughboxes}. The known outer surface constrains the interpretation of radar responses, allowing the geometry of hidden objects to be estimated together with the visible structure. A different approach uses prior knowledge of object shape. In human-body reconstruction, a parametric body template specifies the structure of the mesh, while radar observations provide geometric constraints when image measurements are incomplete. ImmFusion combines image and radar features with body-template information to estimate human joints and meshes under poor lighting or occlusion~\cite{chen2023immfusion}. Radar-based methods have also reconstructed human meshes in real time and through walls~\cite{xue2021mmmesh,zhao2019rfavatar}.

These methods demonstrate how radar measurements can contribute to recovering surfaces beyond direct visual observations, but their applicability depends on the available geometric constraints. Hidden-object reconstruction often assumes a known enclosure and controlled sensing geometry, while human-mesh reconstruction benefits from a strong prior on body structure. Extending these approaches to objects with arbitrary shapes and materials remains difficult, although recent work has reconstructed more general surfaces directly from 3D radar observations~\cite{treffler2026radarsurface}. Another challenge is determining whether the reconstructed geometry actually benefits from wireless measurements. A model may produce plausible meshes primarily from visual information or learned shape priors. Modality-ablation experiments are therefore needed to establish the contribution of radar, and training with randomly omitted modalities can reduce excessive dependence on one sensing source~\cite{chen2023immfusion}.

\subsection{NeRF and 3DGS}
\label{sec:env:radio_aware}

NeRF and 3DGS represent scenes in forms that support image rendering from viewpoints not used during reconstruction. NeRF models spatial density and view-dependent radiance through a continuous neural function, while 3DGS represents scene structure using Gaussian primitives with positions and spatial extents. Both approaches depend on accurate geometric information to reproduce scene appearance from new viewpoints. Wireless measurements can assist these representations by constraining the geometry used during reconstruction. Radar Doppler measurements can further identify moving objects that would otherwise corrupt a static scene representation.

One approach uses RF-derived range information to improve the initialization of the scene representation. Multimodal Gaussian splatting estimates dense depth from sparse range measurements and uses the resulting geometry to initialize Gaussian positions~\cite{gau2026multimodalgs}. Camera observations then optimize the Gaussian representation for visual rendering. In this process, wireless measurements constrain the initial spatial arrangement of the Gaussians, reducing dependence on uncertain monocular depth estimates. The final representation is still optimized using visual observations, so the contribution of RF information depends on the accuracy and spatial distribution of the initial range constraints.

Another approach uses radar motion information to handle moving objects during large-scale mapping. Rad-GS combines a monocular camera with a 4D mmWave radar in a 3DGS-based SLAM framework~\cite{xiao2025radgs}. Radar detections whose Doppler velocities are inconsistent with the estimated ego-motion are labeled as dynamic and projected onto the images to mask moving objects. The remaining radar points, densified with image guidance, are used for pose tracking and as depth constraints on the Gaussian map. Ablation experiments on outdoor driving sequences show that this radar-guided masking improves the rendering quality of the reconstructed scene, and the system has been demonstrated at kilometer scale. In this way, the radial velocity provided by radar sensing enhances the reconstruction by suppressing artifacts caused by moving objects and improving localization accuracy.

The two approaches face different limitations. Range-assisted Gaussian initialization depends on the reliability of sparse RF measurements, and errors in their association with scene structures may affect the subsequent visual optimization. Doppler-based masking relies on accurate ego-motion estimation and radar-camera alignment, and objects moving perpendicular to the radar LoS produce little radial velocity. Radar reflections from transparent surfaces such as glass may also be inconsistent with the visual geometry and cause registration errors. More generally, using RF observations within a scene representation requires accounting for responses that vary with sensing configuration and material properties, which are not directly determined by visual geometry. These issues also arise when a scene representation is expected to support both visual reconstruction and wireless channel prediction, motivating the joint setting discussed in Section~\ref{sec:bridge}.

\section{Vision-Assisted CKM Reconstruction}
\label{sec:chan}

Vision-assisted CKM reconstruction uses visual information about the physical environment to estimate wireless channel quantities at locations or link configurations where direct measurements are unavailable. Visual observations provide the geometry of obstacles and reflecting surfaces, which constrains the propagation paths between a transmitter and a receiver. However, the channel response also depends on the electromagnetic properties of these surfaces, which cannot generally be determined from visual observations alone. Such properties must be assigned through a propagation model or inferred from RF measurements. Existing approaches differ in how they obtain the channel values from the available environmental information. Vision-assisted ray tracing computes propagation paths using a geometric model with electromagnetic parameters. Learning-based approaches predict channel quantities from environmental features and channel labels, while wireless radiance fields fit continuous representations of radio propagation to RF observations. Table~\ref{tab:vision_wireless_recon_taxonomy} summarizes representative methods in these three categories.

\begin{table*}[t]
\centering
\caption{Representative vision-assisted approaches for CKM reconstruction.}
\label{tab:vision_wireless_recon_taxonomy}
\footnotesize
\setlength{\tabcolsep}{3pt}
\renewcommand{\arraystretch}{1.18}
\begin{tabular}{@{}>{\raggedright\arraybackslash}m{0.12\linewidth}
>{\raggedright\arraybackslash}m{0.15\linewidth}
>{\raggedright\arraybackslash}m{0.16\linewidth}
>{\raggedright\arraybackslash}m{0.15\linewidth}
>{\raggedright\arraybackslash}m{0.25\linewidth}
>{\raggedright\arraybackslash}m{0.11\linewidth}@{}}
\toprule
\textbf{Approach} &
\textbf{Representative Work} &
\textbf{Visual Information} &
\textbf{RF Data or Labels} &
\textbf{Role in CKM Reconstruction} &
\textbf{Output} \\
\midrule

Vision-assisted ray tracing &
VisRFTwin~\cite{visrftwin2026} &
Multiview geometry and material estimates &
Sparse channel soundings &
Initializes the scene geometry and electromagnetic parameters for subsequent RF calibration &
Calibrated ray tracer \\

\cmidrule(l){2-6}

&
VLM-guided ray tracing~\cite{kang2026vlmguidedrt} &
Scene images with known geometry &
Simulated RSS samples &
Initializes material conductivities and selects informative calibration positions &
Material conductivities \\

\midrule

Learning-based approaches &
Path-loss radio map~\cite{jaensch2026radiomap} &
Aerial images and surface-height maps &
Ray-tracing labels &
Conditions channel prediction on building layout, height, and transmitter information &
Path-loss map \\

\cmidrule(l){2-6}

&
Point-cloud CKM~\cite{wang2025pointcloudckm} &
Reconstructed 3D point clouds &
Measured channel gains and PDPs &
Selects scene regions consistent with link geometry and propagation delays &
PDP and power maps \\

\cmidrule(l){2-6}

&
Semantic CKM~\cite{wang2025preciseckm} &
3D point clouds with semantic labels &
Measured RSS and PDPs &
Uses geometric and semantic information for link-conditioned channel prediction &
RSS and PDP maps \\

\midrule

Wireless radiance fields &
RFCanvas~\cite{rfcanvas2024} &
Depth, material, and motion cues &
Sparse RF measurements &
Initializes an RF field from visual information and refines its attributes using RF observations &
Editable RF field \\

\cmidrule(l){2-6}

&
RF-3DGS~\cite{rf3dgs2024} &
Visual geometry represented by 3D Gaussians &
Radio spatial spectra &
Fixes the learned visual geometry and fits radio-specific attributes &
Spatial-CSI field \\

\bottomrule
\end{tabular}
\end{table*}

\subsection{Vision-Assisted Ray Tracing}
\label{sec:chan:propagation}

Ray tracing predicts wireless channels by computing propagation paths between a transmitter and a receiver. These paths are determined by the scene geometry, while their channel responses depend on the electromagnetic properties of the interacting surfaces. A visual reconstruction provides the locations and orientations of walls and other objects, allowing the ray tracer to identify possible LoS and reflected paths. However, the reconstructed geometry does not directly provide material parameters such as permittivity and conductivity. These parameters are commonly initialized from material assumptions and subsequently calibrated using channel measurements. The accuracy of the resulting channel predictions therefore depends on both the scene model and the estimated electromagnetic parameters.

Visual information can reduce the RF measurements required for calibration by providing initial estimates of the material properties. For example, VisRFTwin identifies surface materials from multiview images and assigns their corresponding electromagnetic parameters before calibrating the ray tracer with channel soundings~\cite{visrftwin2026}. In its evaluated setting, calibrating the electromagnetic parameters from scratch, without prior material information, yields a prediction error above 8~dB with fewer than 50 measurements, while the error approaches 2~dB only after several hundred measurements. Vision-based initialization improves the starting estimates, allowing accurate calibration with substantially fewer measurements. Visual information can also guide the selection of calibration locations. A vision-language model (VLM)-guided approach selects transmitter-receiver configurations whose propagation paths interact with different materials, making their electromagnetic parameters easier to distinguish from the resulting measurements~\cite{kang2026vlmguidedrt}. The benefit therefore depends not only on the number of RF measurements but also on which surfaces their propagation paths encounter.

Several limitations remain. Materials with similar visual appearances may have different electromagnetic responses because of their composition or surface properties. Consequently, visually assigned parameters still require RF calibration. A surface that is rarely encountered by measured propagation paths may also remain poorly constrained, even when many measurements are available elsewhere in the site. Moreover, an inaccurate scene geometry or antenna model can produce channel errors that are partly absorbed by adjustments to the material parameters. Such compensation may improve the fit to calibration data without improving channel prediction at unmeasured links. Reliable evaluation therefore requires channel measurements from transmitter-receiver configurations that were not used for calibration.

\begin{figure*}[!tb]
    \centering
    \includegraphics[width=0.75\linewidth]{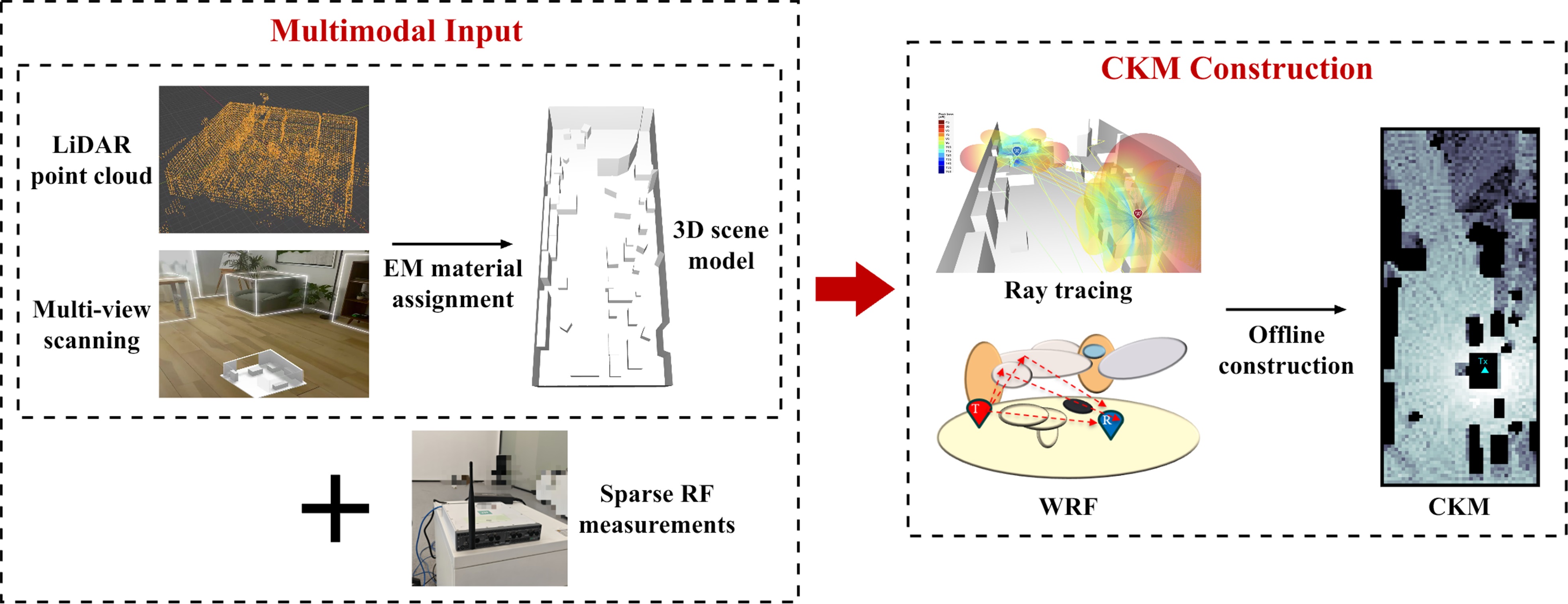}
    \caption{Illustration of vision-assisted CKM reconstruction. LiDAR point clouds and multiview images provide scene geometry and material information for radio propagation modeling. The reconstructed scene can be used by ray tracing to generate a channel-gain map for a specified transmitter configuration. Sparse RF measurements can support model calibration, while wireless radiance fields provide an alternative approach to learning the channel response from visual information and RF observations.}
    \label{forwireless}
\end{figure*}

Figure~\ref{forwireless} illustrates the construction of a specific type of CKM, channel gain map from an indoor scene. LiDAR scanning provides the geometry of the environment, while multiview images can assist in identifying surface materials. After the reconstructed scene is imported into a ray tracer, the transmitter location is specified and the channel gain is computed over a set of receiver positions. The resulting map describes the spatial variation of channel gain for that transmitter configuration. The figure also illustrates how sparse RF measurements may be used to calibrate the propagation model, as well as an alternative route based on wireless radiance fields. These extensions represent general CKM reconstruction approaches and should be distinguished from the ray-tracing example shown here.

\subsection{Learning-Based Approaches}
\label{sec:chan:rmckm}

Learning-based approaches construct CKMs by predicting channel quantities from environmental information and transmitter-receiver configurations. Unlike ray tracing, which explicitly computes propagation paths, these methods learn the relationship between scene features and channel responses from training labels. The labels may be generated by propagation simulators or obtained from field measurements. Visual information provides the spatial structures that affect wireless propagation, allowing a trained model to infer channel values at locations not directly represented in the training observations. The scene information needed by the model depends on the channel quantity being predicted.

For coverage-scale prediction, the spatial distribution of obstacles provides important information about path loss. Aerial images can describe the horizontal layout of buildings, but they do not directly provide building heights, which affect blockage and propagation conditions. A learning-based radio-map approach therefore combines aerial images with surface-height information to predict spatial path-loss distributions~\cite{jaensch2026radiomap}. More detailed channel quantities require finer associations between propagation paths and scene structures. For example, a point-cloud-based method uses the transmitter-receiver geometry to divide the scene into confocal ellipsoidal regions, each corresponding to a time-of-arrival bin of the power delay profile (PDP). The 3D points within each region are selected as structures that may interact with the signals arriving in that bin, and their features are used to predict the channel gain of the corresponding bin~\cite{wang2025pointcloudckm}. Semantic information can further distinguish the types of surfaces represented by these points, providing additional features for channel prediction~\cite{wang2025preciseckm}. These methods illustrate how the required scene representation becomes more detailed when the reconstruction target extends from aggregate channel quantities to delay-resolved information.

The accuracy of learning-based CKM reconstruction depends on the coverage and reliability of the training labels. Simulated labels can be generated for many link configurations, but their predictions inherit the assumptions of the underlying propagation model. Measured labels capture actual channel responses, although collecting them over a large site requires substantial measurement effort. Existing delay-resolved approaches have also been evaluated on limited measured environments, leaving their performance across different sites insufficiently established. Changes in frequency or antenna configuration may alter the relationship between scene features and channel responses. A model trained under one configuration may therefore require additional measurements or adaptation before it can be used elsewhere.

\subsection{Wireless Radiance Fields}
\label{sec:chan:fields}

Wireless radiance fields represent radio propagation through continuous spatial functions learned from RF measurements. Inspired by NeRF and Gaussian-based representations in computer vision, these methods associate spatial locations or directions with radio-related attributes that determine the predicted channel response~\cite{zhao2023nerf2,newrf2024,wrfgs2024}. Unlike conventional ray tracing, the radio attributes can be fitted to measurements without explicitly specifying the electromagnetic parameters of every surface. However, RF observations are usually available for only a limited set of links, and different spatial distributions of radio attributes may explain the same measurements. Visual reconstruction provides geometric constraints that reduce this ambiguity by locating the structures likely to affect propagation.

Existing methods incorporate visual geometry with different degrees of flexibility. RF-3DGS first reconstructs the scene using 3D Gaussians learned from images. It then fixes their positions and shapes while fitting radio-specific attributes to measured spatial spectra~\cite{rf3dgs2024}. Fixing the geometry reduces the number of unknown parameters during RF fitting, but the resulting field remains dependent on the accuracy of the visual reconstruction. RFCanvas instead uses camera-derived information to initialize object geometry and material-related RF attributes, which are subsequently refined using sparse RF measurements~\cite{rfcanvas2024}. When objects move, visually estimated motion updates their positions in the RF field, allowing the model to adapt to geometric changes without reconstructing the entire environment from scratch. These approaches use visual information to constrain the spatial structure of the field, while RF observations determine its radio response.

The treatment of scene geometry introduces a tradeoff. Fixing the visual geometry prevents the RF model from correcting geometric errors that affect propagation, particularly when a surface interacts differently with light and radio waves. Allowing geometry to change during RF fitting provides more flexibility, but the model may explain a channel mismatch by moving a surface when the actual error comes from its radio attributes. The available RF measurements may not be sufficient to distinguish these causes. Furthermore, radio fields fitted under a particular transmitter configuration or frequency do not necessarily generalize to new configurations unless the representation explicitly models the relevant environmental properties. Dynamic scenes introduce an additional difficulty: visual observations can identify geometric changes, but the resulting channel responses still need to be inferred or updated using RF information. These limitations motivate the joint reconstruction approaches discussed in Section~\ref{sec:bridge}.

\section{Toward Joint Physical-Scene and CKM Reconstruction}
\label{sec:bridge}

Sections~\ref{sec:env} and~\ref{sec:chan} have reviewed how wireless observations assist physical-scene reconstruction and how visual information supports CKM reconstruction, respectively. In applications requiring both targets, the two tasks can be jointly performed by exploiting their dependence on the same physical environment. Joint reconstruction aims to recover scene geometry and wireless propagation characteristics from visual and RF observations while maintaining consistency between the two reconstructed targets. Beyond obtaining both outputs, joint reconstruction creates opportunities for mutual refinement, where information from either modality can correct uncertainties in the reconstruction associated with the other.

Such refinement can proceed in both directions. Visual reconstruction identifies the positions of surfaces that may block or reflect radio signals. This geometric information constrains the propagation paths that can explain measured channel responses, thereby supporting CKM reconstruction. Conversely, wireless measurements can provide additional geometric constraints where visual observations are unreliable. For example, a radar range associated with an uncertain surface can refine its estimated position, while resolved multipath delays can provide further constraints on scene geometry, as discussed in Section~\ref{sec:env}. The updated geometry can then be used to improve channel prediction. However, a mismatch between the predicted and measured RF responses does not necessarily indicate incorrect geometry. It may also arise from inaccurate electromagnetic properties. Mutual refinement therefore requires determining which part of the reconstructed environment should be updated from the available observations.

A unified environment representation offers a way to connect the two reconstruction targets. One important principle is to share the underlying scene geometry while accounting for the different responses of optical and RF sensing. The same physical object may interact differently with light and radio waves. For example, cardboard blocks visible light but may allow radar signals to penetrate, depending on the operating frequency. If both modalities are forced to use identical occupancy responses, the model may incorrectly suppress radar returns from objects behind the cardboard. RadarSim addresses this difference by initializing a radar scene field from a camera-trained representation and allowing the radar geometry to deviate from the visual geometry~\cite{chen2026radarsim}. This result illustrates the importance of maintaining modality-specific observation models while exploiting common scene information.

Another important principle is to represent the electromagnetic properties of the environment so that the reconstructed model can be reused under different transmitter-receiver configurations. A wireless field fitted directly to channel measurements from one transmitter may capture the combined effects of the transmitter configuration and the surrounding surfaces. Its predictions may therefore become unreliable when the transmitter changes. If the representation instead characterizes how individual surfaces interact with radio waves, a propagation model can use these properties and the new link configuration to predict channel responses. URF-GS follows this direction by learning scene geometry from images and radio-related surface properties from RF measurements within a Gaussian representation~\cite{urfgs2026}. In simulated indoor environments, the resulting model supports channel prediction under new transmitter-receiver configurations without retraining. This demonstrates the potential of environment-based representations for reusable CKM reconstruction.

Existing studies provide initial evidence for these principles, but fully realizing mutual refinement remains challenging. URF-GS combines visual geometry with learned radio properties, but its geometry is learned from images and kept fixed when the radio properties are fitted. It therefore does not establish a complete process in which RF observations correct a shared scene model and the corrected geometry subsequently improves CKM reconstruction. Such a process also needs to accommodate changes in the environment. When an object moves, visual observations can update its geometry, but the resulting change in radio propagation may affect multiple links across the site. The model must therefore identify which channel predictions require updating and use new RF measurements where the existing environmental information is insufficient, as partly explored in Section~\ref{sec:chan:fields}. Reliable updates also depend on spatial calibration and temporal synchronization between the sensing modalities. Moreover, both reconstruction targets need to be evaluated against corresponding reference observations from the same site, since improved visual reconstruction does not necessarily imply more accurate channel prediction. These requirements motivate further research into cross-modal refinement and consistent environment representations, together with the broader challenges discussed in Section~\ref{sec:future}.

\section{Open Challenges and Future Directions}
\label{sec:future}

Despite the rapid progress reviewed above, multimodal environment reconstruction still faces several challenges before it can be used in practical wireless systems. In this section, we discuss five challenges that are shared by both directions, together with promising directions to address them.

\subsection{Multimodal Datasets and Benchmarks}
\label{sec:future:data}

Although many multimodal datasets combine cameras, LiDAR, and wireless measurements, most of them are collected for specific tasks such as beam prediction or blockage detection~\cite{alkhateeb2023deepsense6g}. Simulated datasets provide richer synchronized modalities and multipath channels, but their channels are generated by ray tracing in virtual scenes~\cite{multimodalwireless2024}. They rarely provide the reference geometry of the scene together with the measured channels at many locations of the same site. As a result, it is difficult to evaluate scene reconstruction and wireless reconstruction on the same site, and to compare different methods fairly. Recent datasets have started to cover low-altitude and indoor scenarios with multiple modalities~\cite{multimodalnf2026,luvira2023}. This motivates shared benchmarks with synchronized visual and RF measurements across diverse sites, frequency bands, and changing scenes, together with metrics that evaluate scene accuracy and channel prediction jointly.

\subsection{Multimodal Calibration and Trustworthy Supervision}
\label{sec:future:calibration}

Cameras, LiDAR, and radio receivers use different coordinate frames, sampling rates, and latencies, so their measurements have to be aligned in space and time before they can be combined. However, even a small alignment error can cause a large error in the reconstruction, because a radio measurement may then be associated with the wrong surface. Moreover, a model may still fit its training data well by adjusting other parameters to compensate for the misalignment, which hides the error. This motivates calibration methods that estimate the sensor poses together with the reconstruction and report the uncertainty of each modality~\cite{multimodalalign2025}.

\subsection{Fidelity and Efficiency Tradeoffs}
\label{sec:future:efficiency}

Practical systems have limited budgets for sensing, computation, latency, and signaling overhead. However, existing methods often reconstruct every part of the scene with the same high fidelity, even though a photorealistic model can be wasteful for channel prediction, and a model tuned for propagation can be visually poor. Therefore, the fidelity of the reconstruction should be allocated according to the target task~\cite{constrainedmultimodal2025}. In particular, since a wireless measurement near an uncertain structure is more valuable than one on a well-reconstructed surface, the sensing resources, such as beams, bandwidth, and measurement positions, can be scheduled to reduce the uncertainty of the reconstruction. This calls for co-designing sensing, communication, and reconstruction~\cite{integratedmultimodal2025}, and for metrics that measure accuracy per unit of sensing and computation cost.

\subsection{Dynamic Scenes and Time-Varying Channels}
\label{sec:future:dynamics}

Most existing methods assume that the scene and the channel are static during reconstruction. This assumption does not hold in vehicular, UAV, and crowded indoor scenarios, where people and vehicles move and the channel can change within milliseconds. Moreover, the scene and the channel change at different rates, so reconstructing the whole model again after every change is too slow and too costly. This motivates representations that separate the static background from moving objects and update only the changed parts, as well as closed-loop sensing and communication systems that update the model continuously~\cite{ainative2025}. Such updating also requires accurate time synchronization, because an observation delay can otherwise appear as object motion.

\subsection{Generalization Across Sites and Configurations}
\label{sec:future:foundation}

Most existing multimodal reconstruction models are trained for a single site, frequency band, and antenna configuration, and they often degrade when any of these changes. Collecting new paired visual and RF data for every site is costly. Large-scale pretraining offers a promising direction, as foundation models for wireless sensing and channel prediction have begun to transfer across scenes and tasks~\cite{wmpm2024,wifo_2024,sansee2024,wifisensinggen2026}. However, such models may produce outputs that look plausible but violate propagation physics, and these errors are hard to detect when reference measurements are scarce. Therefore, embedding propagation laws into the model and pretraining on paired visual and RF data are important steps toward models that generalize to new sites with few measurements.

\section{Conclusion}
\label{sec:conclusion}

This paper reviewed multimodal wireless and visual sensing for environment reconstruction in sixth-generation (6G) environment-aware wireless systems. Wireless measurements improve physical-scene reconstruction by supplying geometry, occupancy, and motion information where visual observations are incomplete or unreliable. Visual geometry, spatial layout, and semantics help construct radio maps and CKMs through ray tracing, learning-based prediction, or wireless radiance fields fitted to sparse RF measurements. Joint reconstruction builds both targets within a unified representation. We discussed design principles for such representations: they should share scene geometry while maintaining modality-specific responses, and they should encode environmental properties for reuse across different transmitter configurations. Existing work demonstrates these principles in single-direction pipelines. The path forward is mutual refinement, where wireless measurements guide scene reconstruction and the refined scene, in turn, improves wireless channel prediction.

\bibliographystyle{IEEEtran}
\bibliography{references}

\end{document}